\documentclass[11pt,a4paper]{article}
\usepackage{jheppub_kim}
\usepackage{rotating} 
\usepackage{graphicx,epsfig}
\usepackage{amsmath}
\usepackage {amssymb}
\usepackage{subfigure}

\usepackage{relsize}

\providecommand{\U}[1]{\protect\rule{.1in}{.1in}}

\newcommand{\newc}{\newcommand}

\newc{\be}{\begin{equation}}
\newc{\ee}{\end{equation}}

\newc{\ba}{\begin{eqnarray}}
\newc{\ea}{\end{eqnarray}}
\newc{\bea}{\begin{eqnarray*}}
\newc{\eea}{\end{eqnarray*}}
\newc{\D}{\partial}
\newc{\ie}{{\it i.e.} }
\newc{\eg}{{\it e.g.} }
\newc{\etc}{{\it etc.} }
\newc{\etal}{{\it et al.}}
\newc{\lcdm}{$\Lambda$CDM }

\newc{\ra}{\Rightarrow}

\title{Modified cosmology through spacetime thermodynamics and Barrow horizon 
entropy}

\author[a,b,c]{Emmanuel N. Saridakis}

\affiliation[a]{National Observatory of Athens, Lofos Nymfon, 11852 Athens, 
Greece}

 \affiliation[b]{Department of Physics, National Technical University of 
Athens, 
Zografou
Campus GR 157 73, Athens, Greece}
\affiliation[c]{Department of Astronomy, School of Physical Sciences, 
University of Science and Technology of China, Hefei 230026, P.R. China}

\emailAdd{msaridak@phys.uoa.gr}

\abstract{ 
We present   modified cosmological scenarios that arise from the application of 
the ``gravity-thermodynamics'' conjecture, using the Barrow entropy instead of 
the usual Bekenstein-Hawking one. The former is a modification of the black 
hole entropy due to   quantum-gravitational effects that deform the black-hole 
horizon by giving it an intricate, fractal structure. We extract modified 
cosmological equations which contain new extra terms that constitute an 
effective dark-energy sector, and which coincide with the usual Friedmann 
equations in the case where the new Barrow exponent acquires its 
Bekenstein-Hawking value. We present analytical expressions for the evolution 
of 
the effective dark energy density parameter, and we show that the universe 
undergoes through the usual     matter and dark-energy epochs. 
Additionally,     
the dark-energy equation-of-state parameter is affected by the value of the 
Barrow deformation exponent and it can lie in the quintessence or phantom 
regime, or experience the phantom-divide crossing. Finally, at asymptotically 
large times the universe always results in the de-Sitter solution.
}

\begin{document}
\maketitle

\section{Introduction}
 
From cosmological observations of various origins we deduce that the universe 
has experienced two phases of accelerated expansion, one at late 
and one at early times. In order to describe this behavior one can either 
construct modified gravitational theories, whose richer structure provides the 
extra degrees of freedom  (for reviews see 
\cite{Nojiri:2006ri,Capozziello:2011et,Cai:2015emx}), or alter the content of 
the universe, by introducing new fields such as the 
inflaton   \cite{Olive:1989nu,Bartolo:2004if} or the dark 
energy  concept \cite{Copeland:2006wr,Cai:2009zp}. Concerning the first 
direction, the  usual approach is to start from the action of general relativity
  and add correction terms, resulting to $f(R)$  gravity
\cite{Starobinsky:1980te,DeFelice:2010aj,Nojiri:2010wj},
  $f(G)$ gravity \cite{Nojiri:2005jg}, Lovelock 
gravity \cite{Lovelock:1971yv}, etc. Alternatively, one 
can construct gravitational modifications using torsion, such as in  
$f(T)$ gravity  
\cite{Ben09,Chen:2010va,Kofinas:2014owa}, non-metricity 
\cite{Iosifidis:2019jgi,Jimenez:2019ovq}, Finsler corrections 
\cite{Basilakos:2013hua} or other geometrical structures.

Additionally, there is a well-known conjecture that gravity and 
thermodynamics are related 
\cite{Jacobson:1995ab,Padmanabhan:2003gd,Padmanabhan:2009vy}, and 
 in   particular one can show that the cosmological Friedmann equations can 
be expressed as the first law of thermodynamics, if we consider the universe as 
a  thermodynamical system bounded by the apparent horizon
\cite{Frolov:2002va,Cai:2005ra,Akbar:2006kj}. Similarly, performing 
the procedure in a reverse way, one can extract the Friedmann equations by 
applying the first law of thermodynamics. The ``gravity-thermodynamics'' 
conjecture is applied very efficiently in a variety of modified theories of 
gravity, with the important step being the use of the 
modified entropy relation which is valid in   each theory 
\cite{Akbar:2006er,Paranjape:2006ca,Sheykhi:2007zp,Jamil:2009eb,
Cai:2009ph,
Wang:2009zv,
Jamil:2010di, Gim:2014nba, Fan:2014ala,Lymperis:2018iuz}.  Hence, although an 
interesting way to 
investigate gravity, as long as  the modified entropy relation is needed, the 
above 
procedure cannot provide new gravitational 
theories, since the modified gravity needs to be known a priori.

Recently, Barrow \cite{Barrow:2020tzx} was inspired by the Covid-19 
virus illustrations and considered the possibility that the 
black-hole surface might have  intricate structure down to arbitrarily small 
scales, due to 
quantum-gravitational effects. Such a fractal structure for the horizon 
leads to finite volume but with infinite (or finite) area.
  Hence, due to the 
basic principle of black hole thermodynamics, the above possible effects of the 
quantum-gravitational spacetime foam on the horizon area will lead to a new 
black hole entropy relation, namely
\begin{equation}
\label{Barrsent}
S_B=  \left (\frac{A}{A_0} \right )^{1+\Delta/2}, 
\end{equation}
where $A$ is the standard horizon area  and $A_0$ the Planck area. The new 
exponent $\Delta$ quantifies the quantum-gravitational deformation, and it is 
bounded as  $0 \leq\Delta \leq 1$, with $\Delta=1$ 
corresponding to   the most intricate and fractal structure, while $\Delta=0$ 
corresponds to the   simplest horizon structure in which case the standard 
Bekenstein-Hawking entropy  is restored  (note that the above formula is similar 
with Tsallis 
nonextensive entropy
\cite{Tsallis:1987eu,Lyra:1998wz,Wilk:1999dr}, although the physical principles 
and interpretation is completely different).
  
In the present manuscript we are interested in applying the 
``gravity-thermodynamics'' 
conjecture in a reverse way, in order to construct new modified gravities, but 
using the Barrow entropy instead of the usual one. In particular, we will 
obtain  modified Friedmann equations, whose extra terms disappear in the case 
where Barrow entropy becomes the standard  Bekenstein-Hawking one.

\section{Modified cosmology through Barrow horizon entropy}
 
Since we are interested in constructing modified Friedmann equations through 
the cosmological application of the ``gravity-thermodynamics'' conjecture, 
using  Barrow entropy, in this section we will first present the basic 
application and then we will extend it using the latter.

\subsection{Friedmann equations from the first law of thermodynamics}
\label{FRWbasicKimcase}
 
We start by presenting the above procedure in the basic case of general 
relativity. We consider  
a  homogeneous and isotropic universe, described by the  
Friedmann-Robertson-Walker (FRW) metric
\begin{equation}
ds^2=-dt^2+a^2(t)\left(\frac{dr^2}{1-kr^2}+r^2d\Omega^2 \right),
\end{equation}
  with $a(t)$ the scale factor, and where $k=0,+1,-1$ corresponds respectively 
to flat, close and open spatial geometry.
 Additionally, we assume that the universe is  filled with 
the matter perfect fluid. According to the ``gravity-thermodynamics'' 
conjecture  the first law can be  interpreted in terms of energy flux through 
local 
Rindler horizons, i.e. it is applied   on the universe horizon 
itself, considered as a thermodynamical system 
separated  by a causality barrier
\cite{Jacobson:1995ab,Padmanabhan:2003gd,Padmanabhan:2009vy}. This horizon is 
generally considered to be the     apparent one 
\cite{Frolov:2002va,Cai:2005ra,Cai:2008gw}
\begin{equation}
\label{FRWapphor}
 {r_{A}}=\frac{1}{\sqrt{H^2+\frac{k}{a^2}}},
\end{equation}
where $H=\frac{\dot a}{a}$ is the Hubble parameter and dots 
representing time-derivatives.

In order to apply the first law of thermodynamics we need to attribute to the 
universe horizon an entropy and a temperature. These are provided by black hole 
thermodynamics, replacing the   black hole 
horizon with the cosmological apparent horizon. For the black-hole temperature  
it is   known that it is inversely proportional to its horizon 
\cite{Gibbons:1977mu}, independently of the underlying gravitational theory, 
and thus for the universe horizon temperature we obtain  
\cite{Padmanabhan:2009vy}
\begin{equation}
\label{Th}
 T_h=\frac{1}{2\pi{r_{A}}}.
\end{equation} 
The black-hole entropy, which as we mentioned  depends on the 
underlying  gravitational theory, in the case of general 
relativity is the usual Bekenstein-Hawking relation $S=A/(4G)$, with $A=4\pi 
r_h^2$   the area of the black hole horizon and $G$ the 
gravitational constant (we use units where $\hbar=k_B = c 
= 1$). Therefore, the apparent horizon entropy is   
\begin{equation}
\label{FRWHorentropy}
S_h=\frac{1}{4G} A.
\end{equation}

In an expanding universe, during a time interval $dt$ the heat 
flow through the horizon  is easily found to be \cite{Cai:2005ra}
 \be \label{FRWenergy}
\delta Q=-dE=A(\rho_m+p_m)H {r_A}dt,
\ee
with  $\rho_m$ and $p_m$ the energy density and pressure of the matter fluid 
that fills the universe. In order to apply the first law of 
thermodynamics   $-dE=TdS$, we need to know $T$ and $dS$. The first is given by 
 (\ref{Th}), while the second is calculated from (\ref{FRWHorentropy}) as 
$dS=2\pi 
\dot{{r}}_A dt/G$, where $\dot{{r}}_A$ can be straightforwardly 
calculated using
(\ref{FRWapphor}). Assembling everything we obtain  
\be
\label{FRWcFE1}
-4\pi G (\rho_m +p_m)= \dot{H} - \frac{k}{a^2}.
\ee 
Finally, imposing  the matter  conservation equation
\be
\label{FRconsrvationWequation}
 \dot{\rho}_m +3H(\rho_m +p_m)=0,
 \ee 
 integration of (\ref{FRWcFE1}) leads to the first Friedmann equation
\be 
\label{FRWcFE2}
\frac{8\pi G}{3}\rho_m =H^2+\frac{k}{a^2}-\frac{\Lambda}{3},
\ee
where the cosmological 
constant arises as an   integration constant. We mention that in the above 
procedure one applies the reasonable equilibrium assumption that the universe 
horizon has the same temperature as the universe fluid, which is true for the 
late-time universe  
\cite{Izquierdo:2005ku,Padmanabhan:2009vy,Frolov:2002va,Cai:2005ra,
Akbar:2006kj,Jamil:2010di}. Lastly, as we mentioned in the 
Introduction, the above 
steps can be extended to modified gravity
theories too, as long as  one uses not the general relativity relation 
  (\ref{FRWHorentropy}), but the modified one of each theory
\cite{Akbar:2006er,Paranjape:2006ca,Sheykhi:2007zp,Jamil:2009eb,
Cai:2009ph,
Wang:2009zv,
Jamil:2010di, Gim:2014nba, Fan:2014ala,Lymperis:2018iuz}.

\subsection{Modified Friedmann equations through Barrow entropy}

We are now ready to apply the procedure the  ``gravity-thermodynamics'' 
conjecture in the case of  Barrow entropy, i.e. extending  the 
procedure described in subsection 
\ref{FRWbasicKimcase}. In particular,  the first law of 
thermodynamics is  $-dE=TdS$, where $-dE$ is still given by 
(\ref{FRWenergy}), $T$ is again given by (\ref{Th}), but now the entropy 
relation will be different, namely it is the Barrow entropy 
 (\ref{Barrsent}) (for the generalized second law of thermodynamics in the 
case of Barrow entropy see \cite{Saridakis:2020cqq}). Hence, we now have  
 \begin{equation}
 dS=(4\pi)^{(1+\Delta/2)} \frac{(2+\Delta)}{ A_{0}^{(1+\Delta/2)} 
 }{r_A}^{\Delta+1}\dot{{r}}_A dt,
\end{equation}
 where we have used that   $A=4\pi 
{r_{A}}^2$. Inserting these relations into the first law of 
thermodynamics, and substituting  $\dot{{r}}_A$ using 
(\ref{FRWapphor}), we finally result to 
\be \label{FRWgfe1}
-(4\pi)^{(1-\Delta/2)}A_{0}^{(1+\Delta/2)}(\rho_m+p_m)=2(2+\Delta) 
\frac{\dot{H}-\frac{k}{a^2}}{\left(H^2+\frac{k}{
a^2}\right)^{\Delta/2}}.
\ee
Lastly, inserting   (\ref{FRconsrvationWequation})  and 
integrating, for the validity region 
$0\leq\Delta\leq1$ we acquire
\be \label{FRWgfe2}
\frac{ (4\pi)^{(1-\Delta/2)}A_{0}^{(1+\Delta/2)} }{6} \rho_m=\frac{2+\Delta 
}{2-\Delta} \left(H^2+\frac{k}{a^2}\right) 
^{1-\Delta/2}-\frac{{C}}{3} A_{0}^{(1+\Delta/2)},
\ee
with ${C}$   the integration constant.

As we can see,  the use of Barrow entropy in the first law of thermodynamics  
resulted to the  modified Friedmann equations 
(\ref{FRWgfe1}) and (\ref{FRWgfe2}), with extra  terms comparing to general 
relativity (note that this is a completely different theory and cosmology 
comparing to the application of Barrow entropy in a holographic context 
\cite{Saridakis:2020zol,Anagnostopoulos:2020ctz}). Focusing on the flat case  
$k=0$ for simplicity, we 
can re-express 
them as 
\begin{eqnarray}
\label{FRWFR1}
&&H^2=\frac{8\pi G}{3}\left(\rho_m+\rho_{DE}\right)\\
&&\dot{H}=-4\pi G \left(\rho_m+p_m+\rho_{DE}+p_{DE}\right),
\label{FRWFR2}
\end{eqnarray}
where
 \begin{eqnarray}
&&
\!\!\!\!\!\!\!\!\!\!\!\!\!\!\!\!\!\!
\rho_{DE}=\frac{3}{8\pi G} 
\left\{ \frac{\Lambda}{3}+H^2\left[1-\frac{ \beta (\Delta+2)}{2-\Delta} 
H^{-\Delta}
\right]
\right\},
\label{FRWrhoDE1}
\end{eqnarray}
\begin{eqnarray}
&& \!\!\!\!\!\!\!\!\!\!\!\!\!\!\!\!\!\!\!\!
p_{DE}= -\frac{1}{8\pi G}\left\{
\Lambda
+2\dot{H}\left[1-\beta\left(1+\frac{\Delta}{2}\right) H^{-\Delta}
\right] 
+3H^2\left[1- \frac{\beta(2+\Delta)}{2-\Delta}H^{-\Delta}
\right]
\right\} 
\label{FRWpDE1}
\end{eqnarray}
respectively are the energy density and pressure of the  effective dark energy 
sector, and with  
$ \Lambda  \equiv 4{C}G(4\pi)^{\Delta/2}$ a parameter with dimensions  
$[L^{-2}]$, and 
$\beta\equiv \frac{4(4\pi)^{\Delta/2}G}{A_{0}^{1+\Delta/2}}$ a parameter with 
dimensions $[L^{-\Delta}]$ (we use units where $\hbar=k_B = c 
= 1$). Hence, the 
effective  
equation of state reads
\begin{eqnarray}
w_{DE}\equiv\frac{p_{DE}}{\rho_{DE}}=-1-
\frac{     
  2\dot{H}\left[1-\beta\left(1+\frac{\Delta}{2}\right) H^{-\Delta}
\right]
 }{\Lambda+3H^2\left[1-\frac{\beta(2+\Delta)}{2-\Delta}H^{-\Delta}
\right]}
\label{FRWwDE}.
\end{eqnarray}
As expected,  in the case $\Delta=0$   the modified 
Friedmann equations  (\ref{FRWFR1}),(\ref{FRWFR2}) reduce to $\Lambda$CDM 
paradigm, i.e. 
\begin{eqnarray}
&&H^2=\frac{8\pi G}{3} \rho_m+\frac{\Lambda}{3}\nonumber\\
&&\dot{H}=-4\pi G(\rho_m+p_m).
\end{eqnarray}

\section{Cosmological implications}
\label{CosmEvol}

In the previous section we applied the ``gravity-thermodynamics'' conjecture 
with the Barrow entropy and we resulted in a modified cosmology, characterized 
by the modified Friedmann equations  (\ref{FRWFR1}) and (\ref{FRWFR2}). These 
equations coincide with $\Lambda$CDM paradigm in the limit $\Delta=0$, in which 
case Barrow entropy becomes the standard one, however in the general case they 
give rise to an effective dark energy sector. In the present section we will 
investigate the cosmological evolution, extracting analytical solutions.   

We focus on the case of dust matter ($ p_m\approx0$), in which case as usual  
(\ref{FRconsrvationWequation}) leads to  $\rho_{m} = \frac{\rho_{m0}}{a^3}$, 
with $\rho_{m0}$ the  matter energy density at the current scale factor $a_0=1$ 
(from now on the subscript ``0" denotes the   value of a quantity at present 
time). Furthermore, we introduce the  density 
parameters as 
 \begin{eqnarray} \label{FRWomatter}
&&\Omega_m=\frac{8\pi G}{3H^2} \rho_m\\
&& \label{FRWode}
\Omega_{DE}=\frac{8\pi G}{3H^2} \rho_{DE},
 \end{eqnarray} 
for the matter and effective dark energy sector respectively.
 Hence,  (\ref{FRWomatter}) gives
$\Omega_m=\Omega_
{m0} H_{0}
^2/a^3 H^2$, which  knowing that  $\Omega_m + 
\Omega_{DE}=1$  leads to
\be \label{h2}
H=\frac{\sqrt{\Omega_{m0}} H_{0}}{\sqrt{a^3 (1-\Omega_{DE})}}.
\ee

It proves convenient to  use the redshift  $ 
 1+z=1/a$ as the independent variable. Differentiating (\ref{h2}) we acquire
\be \label{FRWhddot}
\dot H=-\frac{H^2}{2(1-\Omega_{DE})}[3(1-\Omega_{DE})+(1+z)\Omega'_{DE}],
\ee
where   primes mark $z$-derivatives. Inserting (\ref{FRWrhoDE1}) into 
(\ref{FRWode}) and using (\ref{h2}) and  (\ref{FRWhddot}) we acquire a simple 
differential equation for $\Omega_{DE}(z)$, which can be easily solved as
 \begin{eqnarray} 
 \label{omegaDE}
\Omega_{DE}(z)=
1-H^{2}_{0}\Omega_{m0}(1+z)^3 
 \left\{\frac{
(2\!-\!\Delta)}{\beta 
(2+\Delta)}\left[H^{2}_{0}\Omega_{m0}(1\!+\!z)^3+\frac{\Lambda}{3}
\right]\right\}^{\frac{2}{\Delta -2}}\!.
 \end{eqnarray}
 This is the  analytical solution for the dark energy density 
parameter, in the case of dust matter in a flat universe. Lastly, applying it 
at   $z=0$  we obtain
\begin{eqnarray}
\label{lambda}
 \Lambda=\frac{3\beta(2+\Delta)}{2-\Delta}H_0^{(2-\Delta)}-3H_0^2\Omega_{m0},
\end{eqnarray}
 which  leaves the scenario with two 
free 
parameters since it can be used to eliminate one of $\Lambda$, $\Delta$ and 
$\beta$ in terms of the observationally 
determined quantities $\Omega_{m0}$ and $H_0$. We mention that 
   for $\Delta=0$    we re-obtain $\Lambda$CDM concordance scenario. 

Concerning the important observable quantity, the dark-energy 
equation-of-state    parameter given in  (\ref{FRWwDE}), inserting $\dot{H}$ 
from (\ref{FRWhddot}) we find that
\be
\label{wDEfinal}
w_{DE}(z)=-1+\frac{\left\{3[1-\Omega_{DE}(z)]+(1+z)\Omega_{DE}'(z)\right\}
\left\{1-\beta 
(1+\Delta/2) 
\left[\frac{
H^{2}_{0}\Omega_{m0}(1+z)^3}{1-\Omega_{DE}(z)}\right]^{-\Delta/2}\right\}}{[
1-\Omega_{DE}
(z)]\left\{\frac{\Lambda 
[1-\Omega_{DE}(z)]}{H^{2}_{0}\Omega_{m0}(1+z)^3}+3\left\{1-\frac{\beta 
(2+\Delta)}{2-\Delta}\left[\frac{H^{2}_{0}\Omega_{m0}(1+z)^3}{1-\Omega_{DE}(z)}
\right]^{-\Delta/2}\right\} \right\}},
\ee
where $\Omega_{DE}'(z)$ is calculated from 
 (\ref{omegaDE}) as
 \begin{eqnarray} 
\label{omegaDEdot}
&&\!\!\!\!\!\!\!\!\!\!\!\!\!\!\!\!\!\!\!\!\!
\Omega'_{DE}(z)=\left\{ 
\frac{(2-\Delta)}{\beta(2+\Delta)}\left[1+\frac{\Lambda}{3}\frac{1}{\Omega_{m0}
H^{2}_{0}(
1+z)^3}\right]\right\}^{\frac{4-\Delta}{\Delta -2}}
\nonumber\\
&& \ \ \, \   \ 
\cdot
\frac{1}{\beta 
(2+\Delta)}\left[\Omega_{m0}H^{2}_{
0}(1+z)^3\right]^\frac{2}{\Delta -2}
\left[3\Delta \Omega_{m0}H^{2}_{0}(1+z)^2+(\Delta 
-2)\frac{\Lambda}{1+z}\right].
 \end{eqnarray} 
 In conclusion, we were able to extract analytical 
solutions for the effective dark energy density and its equation of state, for 
the modified cosmology arisen from Barrow entropy. 

\begin{figure}[!h]
\centering
\includegraphics[width=7.25cm]{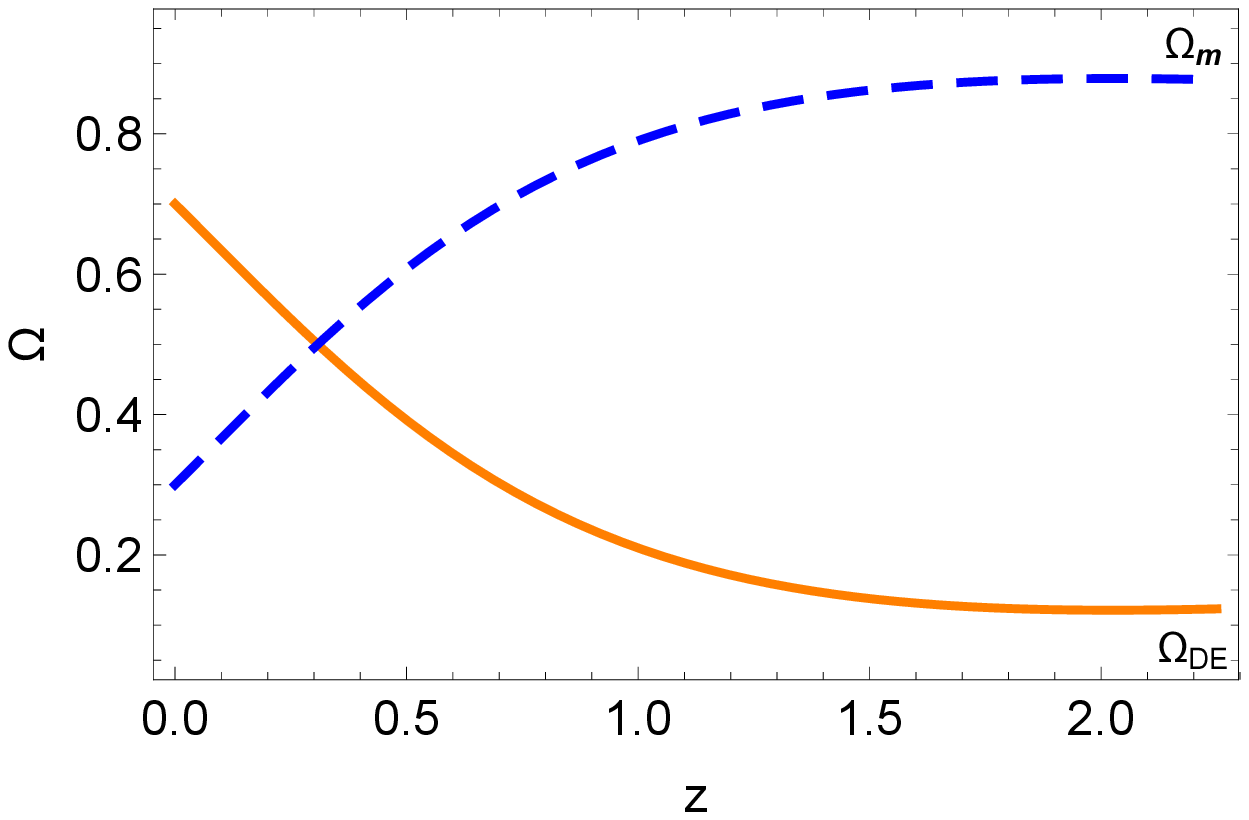}                                    
\includegraphics[width=7.25cm]{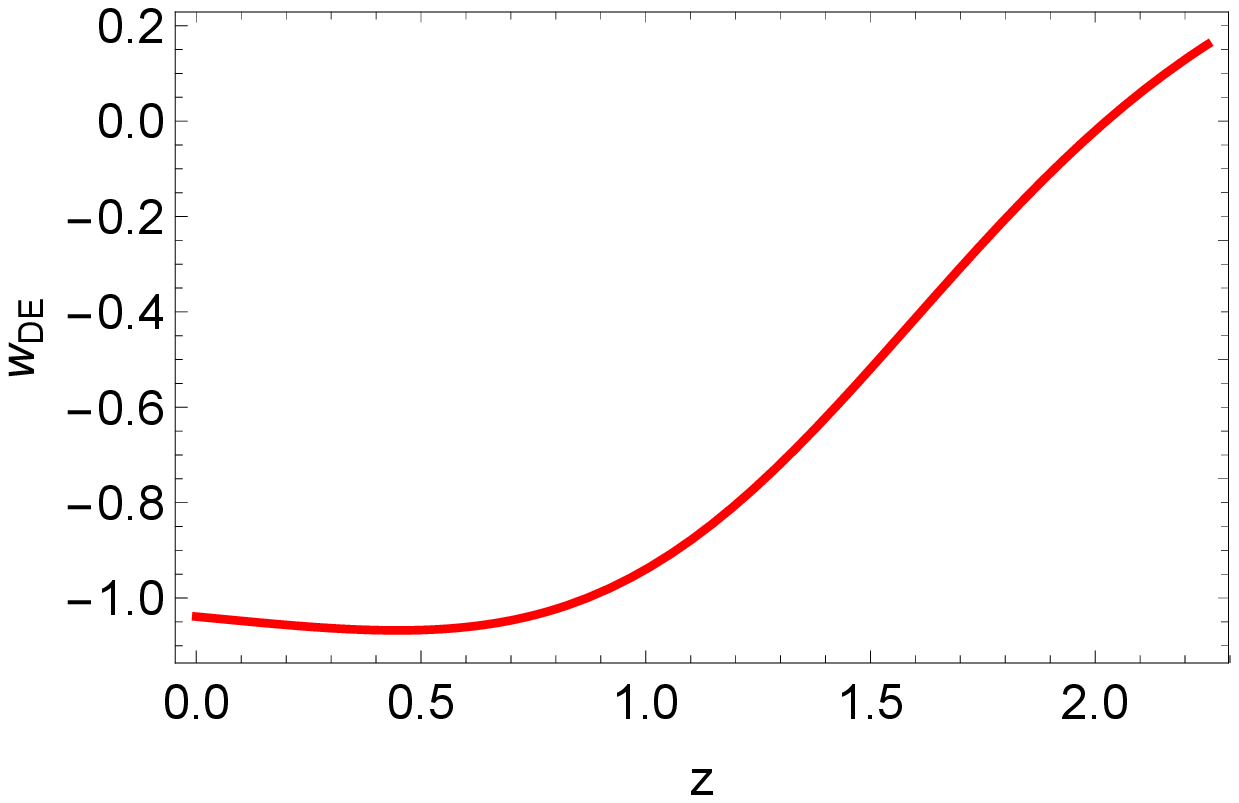} 
\caption{\it{Left graph: The evolution of the effective dark energy density 
parameter $\Omega_{DE}$ (orange-solid) and of the matter density parameter 
$\Omega_{m}$ 
(blue-dashed),   as a function of the redshift $z$, for the 
modified 
cosmology through Barrow entropy with $\Delta=0.2$ and $A_0=1$. Right graph: 
The evolution of the corresponding dark-energy 
equation-of-state parameter
$w_{DE}$.  We have imposed $ \Omega_{m0}\approx0.3$ at 
present time. 
}}
\label{Omegas}
\end{figure}

Let us examine the scenario in more  detail. We start with the case where 
   the explicit cosmological constant $\Lambda\neq0$. Since for 
  $\Delta=0$  we re-obtain 
$\Lambda$CDM paradigm,   we are interested in investigating the role of the 
new Barrow parameter $\Delta$ on the  universe evolution.
We use relation (\ref{lambda}) in order to eliminate   $\Lambda$, imposing the 
observed value
$\Omega_{m0}\approx0.3$   \cite{Ade:2015xua}, and we set $A_0=1$. Using 
(\ref{omegaDE}), in the left graph of Fig.  \ref{Omegas} we present 
$\Omega_{DE}(z)$ and $\Omega_{m}(z) = 
1-\Omega_{DE}(z)$, for  the case where $\Delta=0.2$. Additionally, in the right 
  graph we depict the   evolution of the dark-energy equation-of-state 
parameter  from  (\ref{wDEfinal}).  
As we can see, we are able to obtain the thermal history of the universe in 
agreement with observations. Moreover, the dark energy equation-of-state 
parameter can experience the phantom-divide crossing, which is an advantage of 
the scenario. 
 
\begin{figure}[!h]
\centering
\includegraphics[width=9.5cm]{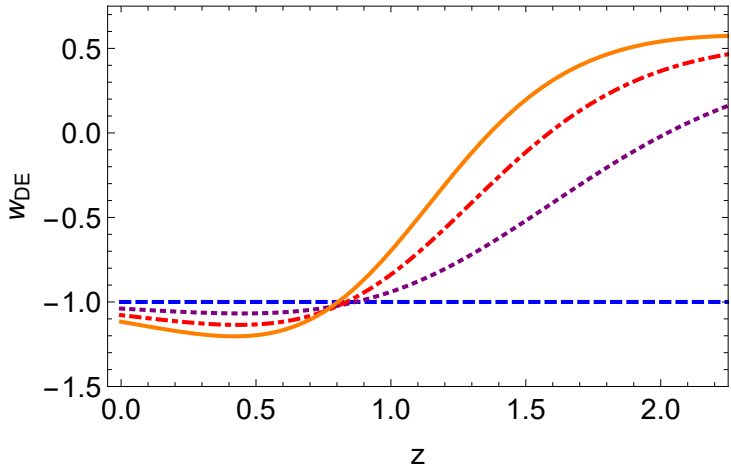}
\caption{\it{The evolution of  
$w_{DE}$ as a function of the redshift $z$, for $A_0=1$,
 and  
for 
 $\Delta=0$ (blue-dashed), $\Delta=0.2 $ 
(purple-dotted), $\Delta=0.4$ 
(red-dashed-dotted), and   $\Delta=0.6$ (orange-solid). We have imposed 
$ \Omega_{m0}\approx0.3$ at 
present time.}}
\label{figwDE}
\end{figure}

In order to examine the effect of the Barrow parameter $\Delta$ on the dark 
energy features,  in Fig.  \ref{figwDE} we 
present $w_{DE}(z)$   for various values of $\Delta$.
 As expected,   for $\Delta=0$ we  acquire  $w_{DE}=-1=const.$, namely 
$\Lambda$CDM scenario. Nevertheless, as the Barrow exponent increases, and the 
quantum-gravitational deformation becomes more important, $w_{DE}$ at larger 
redshifts acquires larger values, while at small redshifts and current time
it  acquires algebraically  smaller values.  Hence, the Barrow parameter 
  $\Delta$, that lies in the core of the modified cosmology at hand, leads the 
dark energy to have  a dynamical nature, departing from  $\Lambda$CDM 
cosmology. Furthermore, dark energy can be  quintessence-like, phantom-like, or 
experience the phantom-divide crossing during the evolution, which is an 
advantage of the scenario.

Finally, we can calculate analytically the asymptotic value of  $w_{DE}$ in the 
far future. In particular, taking the limit $z \rightarrow -1$ in 
(\ref{omegaDE}),(\ref{omegaDEdot}) and  
(\ref{wDEfinal}), we respectively find   $\Omega_{DE}\rightarrow 1$, 
$\Omega'_{DE}\rightarrow 0$, 
and  $w_{DE}\rightarrow-1$. This implies that although at intermediate times
  the dark-energy equation-of-state parameter experiences an interesting 
behavior which departs from $\Lambda$CDM cosmology,   at asymptotically large 
times it will always stabilize at the cosmological constant value $-1$, and the 
universe will reach the  de-Sitter solution, independently of the Barrow 
exponent. The fact that the de Sitter solution is   a stable late-time 
attractor  independently of the Barrow 
exponent,   is a significant advantage  of the modified cosmology through 
Barrow entropy.

We close this analysis by investigating the interesting case where an explicit 
cosmological constant $\Lambda$ is absent, and thus the modified cosmology at 
hand offers a more radical modification, without possessing  $\Lambda$CDM 
paradigm as a particular limit,  which however can still describe the effective 
dark energy and  late-time acceleration. In the case  $\Lambda=0$, relations 
(\ref{FRWrhoDE1}), (\ref{FRWpDE1}) become
\begin{eqnarray}
&& \!\!\!\!\!\!\!\!\!\!\!
\rho_{DE}=\frac{3}{8\pi G}\left[H^2-\Omega_{m0}  (H_{0}/H)^{\Delta}
\right]
\label{rhoDE3}\\
&& \!\!\!\!\!\!\!\!\!\!\!
p_{DE}= -\frac{1}{8\pi G}\left\{
3H^2\left[1-\Omega_{m0}(H_{0}/H)^{\Delta}
\right]
 +2\dot{H}\left[1-\Omega_{m0} (1-\Delta/2) (H_{0}/H)^{\Delta}
\right]
\right\}.
\label{pDE3}
\end{eqnarray}
Hence,  (\ref{omegaDE}) is now written as 
\be
 \label{FRWomegaDE1b}
\Omega_{DE}(z)=1-\Omega_{m0}(1+z)^{\frac{3\Delta}{\Delta-2}}.
\ee 
and   (\ref{wDEfinal})  as
\be \label{wDEfinal1}
w_{DE}(z)=\frac{\Delta}{(2-\Delta)} 
 \left[1-\Omega_{m0}(1+z)^\frac{3\Delta}{(\Delta -
2)} 
\right]^{-1}.
\ee
 Notice that in this case   $\Lambda$CDM scenario cannot be re-obtained 
for any parameter values, and therefore  one should use a non-trivial value for 
$\Delta$ in order to suitably acquire agreement with observations. 
Finally, we mention that from  (\ref{FRWomegaDE1b}) we observe that we are able 
to obtain the usual thermal history of the universe, with the sequence of 
matter and dark energy eras, while  
in the asymptotically far future  ($z \rightarrow -1$) the universe tends 
to the complete 
dark-energy domination.  Lastly, note that  according to (\ref{FRWomegaDE1b}), 
for high redshifts we obtain   either early-time dark 
energy or   $\Omega_{DE}(z)<0$, cases that are not physically interesting. 
However, as expected, these are eliminated if one includes the radiation sector 
too, which changes and regulates the  early-time  behavior.

\section{Conclusions}
\label{Conclusion}

There is a long-standing  conjecture that gravity is related to 
thermodynamics, which concerning cosmological frameworks implies that the 
Friedmann   equations can arise from the first law of thermodynamics.
In this manuscript we constructed   modified cosmological scenarios through the 
application of this conjecture, but using the 
 Barrow entropy, instead of the usual   Bekenstein-Hawking 
one. In particular, as it was recently proposed in 
\cite{Barrow:2020tzx},   the
black-hole surface may have  intricate, fractal structure, due to 
quantum-gravitational effects. Hence, the corresponding black-hole entropy will 
deviate from the Bekenstein-Hawking one, and this deformation is quantified 
through a new exponent $\Delta$, with the limit $\Delta=0$ corresponding to the 
standard case where Barrow entropy becomes the standard one,  while the limit  
$\Delta=1$ corresponds to the case where the deformation is maximal.

Applying the ``gravity-thermodynamics'' procedure with Barrow entropy we 
resulted to modified cosmological equations which contain new extra terms, and 
which coincide with the usual Friedmann equations in the case where the new 
Barrow exponent acquires its usual value $\Delta=0$. In the general case
these new terms constitute an effective dark energy sector leading to 
interesting phenomenological behavior, while in the special case  $\Delta=0$ 
 $\Lambda$CDM concordance model is restored.   

Assuming the matter sector to be dust, we extracted analytical expressions 
for the evolution of the effective dark energy density parameter and its  
equation of state. As we saw, the scenario at hand can describe  the usual 
thermal history of the universe, with  the dark-energy epoch following the 
matter one. Concerning the dark-energy equation-of-state parameter we saw that 
in the recent and current universe it is affected by the value of the Barrow 
deformation exponent. Specifically,   $w_{DE}$ at larger 
redshifts acquires larger values, while at small redshifts and current time
it  acquires algebraically  smaller values, and moreover dark energy can be  
quintessence-like, phantom-like, or 
experience the phantom-divide crossing during the evolution. However,    at 
asymptotically large 
times  the universe will reach the  de-Sitter solution, independently of the 
Barrow exponent, which is an additional advantage. Finally, the scenario at 
hand exhibits interesting cosmological behavior even in the case where an 
explicit cosmological constant is absent, where the modification is more 
radical and the effective dark-energy sector is constituted  solely by the new
terms.

In conclusion,  modified cosmology through ``gravity-thermodynamics'' procedure 
using Barrow entropy  is efficient in quantifying the universe evolution in 
agreement with observations. It would be interesting to perform a full 
observational confrontation using data from  Supernovae type Ia
data  (SNIa),  Cosmic Microwave Background (CMB) shift parameters, 
Baryonic Acoustic Oscillations (BAO),  growth rate and Hubble data 
observations,  in order to  constrain the new Barrow exponent $\Delta$. This 
works lies beyond the scope of the present work and it is left for a future 
project.


\providecommand{\href}[2]{#2}\begingroup\raggedright\endgroup
\end{document}